# Single-photon quantum error rejection and correction with linear optics

Demetrios Kalamidas
*Institute for Ultrafast Spectroscopy and Lasers, City College of the City University of New York,
138 Street & Convent Ave, New York, NY, 10031, USA*

We present single-photon schemes for quantum error rejection and correction with linear optics. In stark contrast with other known proposals, our schemes do not require multi-photon entangled states, are not probabilistic, and their application is not restricted to single bit-flip errors.

The existence of theoretical protocols that enable the detection and correction of errors on quantum states [1-4] has engendered hope with regard to the feasibility of long-distance quantum communication with photons [5,6]. However, because coupling between photons is extremely weak, the realization of the two-qubit gates [7] required in the theoretical protocols would be exceedingly difficult and inefficient. Fortunately, it is possible to partially implement certain two-qubit gate operations with the use of linear-optical elements. So far, the linear-optical approach to quantum error detection and correction is based on the parity check [8,9]. In this Letter we present two linear-optical schemes that, instead of relying on parity checks between two photon qubits, borrow an idea from 'time-bin entanglement' [10] and apply it to the encoding and decoding of single-photon qubits for the purpose of rejecting and correcting errors. Although only single-photon qubits are required to implement the schemes, a practical way of obtaining them is from photon pairs produced by pulse-pumped *spontaneous parametric down-conversion* (SPDC) [11], where one photon serves as a trigger and the other photon is encoded. For clarity and concise exposition of the working principles of the schemes, we assume ideal optical components and do not consider photon loss [6].

Consider the scheme of Fig.1. Alice has a single-photon qubit in an arbitrary unknown state $|\psi\rangle = \alpha|H\rangle + \beta|V\rangle$, where the kets denote the 'horizontal' and 'vertical' polarization modes of the photon, respectively, and $|\alpha|^2 + |\beta|^2 = 1$. She wants to transmit $|\psi\rangle$ to Bob over a noisy channel in a manner that enables him to reject any qubit errors that may have occurred and keep only the uncorrupted states. To this end, she possesses an unbalanced polarization interferometer, based upon two polarizing beam splitters

(PBS), and a fast Pockels cell (PC). The PBS transmit $|H\rangle$ photons (so that they propagate through the short path, S) and reflect $|V\rangle$ photons (so that they propagate through the long path, L). If Alice's photon is in the form of an ultrashort wave-packet (~100fs) typical of pulse-pumped SPDC, and the time-of-flight difference of the unbalanced interferometer is on the order of a few nanoseconds, her qubit transforms as $|\psi\rangle \to \alpha|H\rangle_S + \beta|V\rangle_L$. Alice activates PC$_A$ only when the L-path component is present, effecting the transformation $|V\rangle_L \to |H\rangle_L$. Hence, the final encoded state launched into the noisy channel is of the form $\alpha|H\rangle_S + \beta|H\rangle_L$.

Suppose that the noisy channel is a long-distance optical fiber, where random birefringence (essentially due to thermal fluctuations, vibrations, and imperfections of the

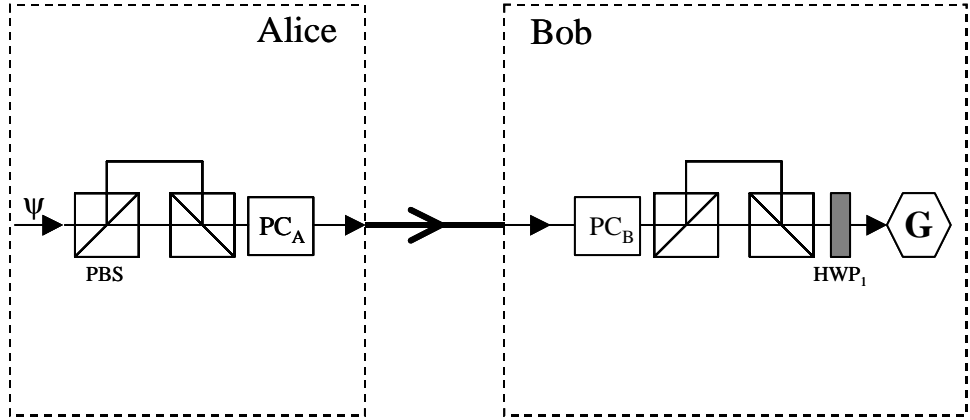

Fig.1: Scheme for single-photon quantum error-rejection. Alice encodes her qubit in two time-bins and sends it to Bob over a noisy channel. Bob decodes the qubit he receives so that the uncorrupted qubit emerges only at specific times during the transmission protocol while projections onto error-states emerge at different times.

fiber itself) induces unknown transformations of the polarization state. However, because Alice has assigned the components of her single-photon qubit to two different time-bins, a few nanoseconds apart, the above causes of birefringence are virtually in a steady-state during this temporal interval. In other words, with this encoding technique, whatever unknown unitary operator, U, acts on the early component also acts on the later component. The operator U can be expressed by the transformation $|H\rangle \to e^{i\phi}\cos\theta|H\rangle + e^{i\chi}\sin\theta|V\rangle, |V\rangle \to -e^{-i\chi}\sin\theta|H\rangle + e^{-i\phi}\cos\theta|V\rangle$ which describes a

general qubit transformation (excluding a global phase which is of no physical significance in this context). However, because of Alice's encoding, the polarization state in both time-bins is $|H\rangle$ and the action of U over the long-distance channel can be described by $|H\rangle_j \to e^{i\phi}\cos\theta|H\rangle_j + e^{i\chi}\sin\theta|V\rangle_j$, where the subscript j=S,L denotes the time-bin. The evolution of $|\psi\rangle$ from Alice to Bob can be written as

$$|\psi\rangle \xrightarrow{Interferometer} \alpha|H\rangle_S + \beta|V\rangle_L \xrightarrow{PC_A} \alpha|H\rangle_S + \beta|H\rangle_L \xrightarrow{U}$$

$$\alpha(e^{i\phi}\cos\theta|H\rangle_S + e^{i\chi}\sin\theta|V\rangle_S) + \beta(e^{i\phi}\cos\theta|H\rangle_L + e^{i\chi}\sin\theta|V\rangle_L) = |\psi\rangle_U. \quad (1)$$

Bob is equipped with $PC_B$, the same polarization interferometer as Alice, a half-wave plate ($HWP_1$), and a time-gate (G). Bob activates $PC_B$ only when the S-path components of the qubit are present, effecting the transformation $|H\rangle_S \leftrightarrow |V\rangle_S$. $HWP_1$ effects the transformation $|H\rangle \leftrightarrow |V\rangle$. The action of $PC_B$, the interferometer, and $HWP_1$ on the received state $|\psi\rangle_U$ is given by

$$|\psi\rangle_U \xrightarrow{PC_B} \alpha(e^{i\phi}\cos\theta|V\rangle_S + e^{i\chi}\sin\theta|H\rangle_S) + \beta(e^{i\phi}\cos\theta|H\rangle_L + e^{i\chi}\sin\theta|V\rangle_L) \xrightarrow{Interferometer}$$

$$\alpha(e^{i\phi}\cos\theta|V\rangle_{SL} + e^{i\chi}\sin\theta|H\rangle_{SS}) + \beta(e^{i\phi}\cos\theta|H\rangle_{LS} + e^{i\chi}\sin\theta|V\rangle_{LL}) \xrightarrow{HWP_1}$$

$$e^{i\phi}\cos\theta(\alpha|H\rangle_{SL} + \beta|V\rangle_{LS}) + e^{i\chi}\sin\theta(\beta|H\rangle_{LL} + \alpha|V\rangle_{SS}). \quad (2)$$

From the last line of expression (2) we see that the first term indicates that the original qubit state (free of errors) emerges at a definite time-of-arrival, corresponding to the delay of propagation once through path S and once through path L (SL or LS). The second term indicates that the $|H\rangle$ and $|V\rangle$ components of the qubit are temporally separated and arrive too late (LL) or too early (SS), respectively. Therefore, Bob's time gate (which can simply be a computer with time-tagging software that is connected to a detector) can readily discard all events that correspond to the transmitted qubit having been projected onto an error-state. With this encoding/decoding technique, the parameters ϕ and χ cannot induce errors and the uncorrupted qubit state is obtained with a probability equal to $\cos^2\theta$. This property is desirable since it means that for small values of $\theta$ the probability is close to 1. Allowing $\theta$ to vary over its entire range during transmissions (indicating strong environmental influence on the channel) only means that the probability of obtaining the uncorrupted state tends to $\frac{1}{2}$.

The protocol of Fig.1 has significant advantages over other proposed linear-optical schemes for quantum error rejection [12,13], including a recent experimental demonstration [14]. The alternative schemes only work for bit-flip errors while our method rejects *any* qubit error that may occur due to the action of an unknown unitary operator over the noisy channel. Because the alternative schemes rely on parity checks within multi-photon entangled states, a bit-flip on two qubits renders the method ineffective. This constrains the bit-flip probability to values much less than 1 so that fatal double-errors become negligible. In our case there is only a single-photon qubit involved in the process and therefore the 'double-error' possibility does not exist, there is no need of multi-photon entanglement, and the variation of the error-inducing parameter, $\theta$, does not affect the error-rejecting capability of the scheme. Furthermore, because of the linear-optical encoding of the unknown single-photon qubit within a multi-photon entangled state, the alternative schemes are inherently probabilistic and require post-selection in the coincidence basis in order to determine whether proper encoding was achieved. The scheme of Fig.1 is deterministic because every encoding attempt on the unknown qubit state (by Alice) is valid and every decoding attempt (by Bob) succeeds in revealing an error on the transmitted qubit state.

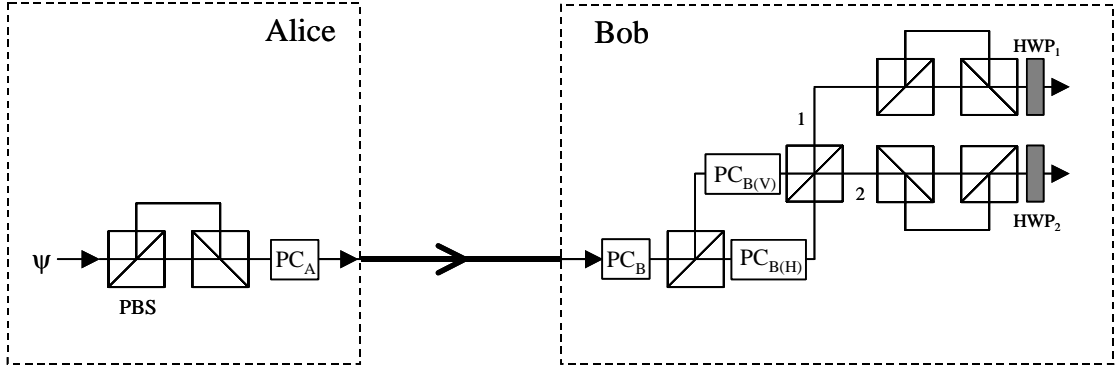

Fig.2: Scheme for single-photon quantum error-correction. Alice and Bob employ Pockels cells so that the method for error-rejection can be extended to error-correction. After his decoding, Bob *always* obtains the uncorrupted qubit sent by Alice, in one or the other of his output channels, at definite times during the transmission protocol.

We can now demonstrate how the basic idea of the previous error rejection scheme may be incorporated into a more complex system that enables error correction on the transmitted single-photon state. The scheme for error correction is depicted in Fig.2. Once more, fast Pockels cells will play a central role in the encoding and decoding of the qubit. As before, the function of each PC, when activated, will be to effect the $|H\rangle \leftrightarrow |V\rangle$ transformation only at specific times during the transmission protocol. Alice encodes her qubit state in the manner of Fig.1 and the state received by Bob is $|\psi\rangle_U$ as given by (1). Bob activates $PC_B$ only when the S-path components of $|\psi\rangle_U$ are present. The state then propagates through a balanced polarization interferometer where $PC_{B(H)}$ is activated only when the S-path component is present and $PC_{B(V)}$ is activated only when the L-path component is present. At each output, 1 and 2, of this interferometer there is another polarization interferometer, identical to Alice's, followed by a HWP effecting the $|H\rangle \leftrightarrow |V\rangle$ transformation. The evolution of $|\psi\rangle_U$ through Bob's configuration can be written as

$$|\psi\rangle_U \xrightarrow{PC_B} \alpha(e^{i\phi}\cos\theta|V\rangle_S + e^{i\chi}\sin\theta|H\rangle_S) + \beta(e^{i\phi}\cos\theta|H\rangle_L + e^{i\chi}\sin\theta|V\rangle_L) \xrightarrow[PC_{B(V)}]{PC_{B(H)}}$$

$$\alpha(e^{i\phi}\cos\theta|V\rangle_S^1 + e^{i\chi}\sin\theta|V\rangle_S^2) + \beta(e^{i\phi}\cos\theta|H\rangle_L^1 + e^{i\chi}\sin\theta|H\rangle_L^2) \xrightarrow[HWP_1, HWP_2]{Interferometers}$$

$$\alpha(e^{i\phi}\cos\theta|H\rangle_{SL}^{1} + e^{i\chi}\sin\theta|H\rangle_{SL}^{2}) + \beta(e^{i\phi}\cos\theta|V\rangle_{LS}^{1} + e^{i\chi}\sin\theta|V\rangle_{LS}^{2}) =$$
$$e^{i\phi}\cos\theta(\alpha|H\rangle_{SL}^{1} + \beta|V\rangle_{LS}^{1}) + e^{i\chi}\sin\theta(\alpha|H\rangle_{SL}^{2} + \beta|V\rangle_{LS}^{2}). \qquad (3)$$

In expression (3) the ket superscripts, 1 and 2, denote the output mode from the balanced interferometer. From the last line of (3) we see that Bob *always* obtains the uncorrupted qubit state at a definite time-of-arrival, albeit in two output modes, despite its transmission through a noisy channel. As before, if $\theta$ takes on small values, Bob obtains the uncorrupted state in mode 1 with high probability and, under this condition, mode 2 can be regarded as the 'error-correcting' mode. If $\theta$ is allowed to vary over its entire range then the probability of obtaining the uncorrupted state in either mode tends to $\frac{1}{2}$. Interestingly, the scheme of Fig.2 exhibits a unique 'self-correcting' property: For every qubit transmission, just applying the protocol ensures that the uncorrupted qubit state is obtained by Bob, without the need of first 'becoming aware' of an error and subsequently applying the corrective action. This property in itself is important because it enables one to avoid experimentally challenging features such as fast feed-forward control and quantum data storage. Again, because the method requires only single-photon qubits, it has the same significant advantages as those mentioned for the scheme of Fig.1: Other proposals for quantum optical error correction [15], including a recent experimental demonstration [16], rely on multi-photon entanglement and parity checks, are probabilistic, and can only deal with single bit-flip errors.

In summary, we have presented single-photon linear-optical schemes for quantum error rejection and correction. The first scheme enables the rejection of any error on the received qubit state. The second scheme is 'self-correcting' and, in principle, enables every transmitted qubit to be obtained in an uncorrupted state. However, each received qubit emerges randomly in either one of the two output modes according to a distribution that depends on the variation of the relevant error parameter. This implies a redundancy in the 'applications package' that will further process the received qubits if all of them are to be utilized. Both schemes include local interferometers that must be kept stable in order to ensure the proper functioning of the protocols. This task is non-trivial and delicate but has been demonstrated many times in other quantum optics demonstrations of quantum information protocols (such as teleportation, entanglement purification, and cryptography) [5]. More importantly, despite the fact that our two schemes represent dramatic simplifications with regard to performing linear-optical quantum error rejection and correction, they possess major advantages over the more complicated alternative schemes. To our knowledge, all other proposals for linear-optical realizations of quantum error rejection and correction rely on parity checks within multi-photon entangled states, are inherently probabilistic, have very low bit rates, and can only deal with single bit-flip errors. In this light we believe that, because of their experimental feasibility and good performance, the schemes described in this Letter may be of considerable use when trying to implement various quantum communication protocols.

I am grateful to D.M. Greenberger and B. B. Das for their encouragement. Partial financial support for this research was provided by DoD.